\UseRawInputEncoding
\documentclass[preprint,showkeys,aps,prb]{revtex4}
\usepackage{amsmath}
\usepackage[dvips]{graphicx}
\usepackage{setspace}

\begin{document}


\title{
Time-Symmetry Breaking in Hamiltonian Mechanics. III. \\
A Memoir for Douglas James Henderson [1934-2020]      \\
}

 \author{
William Graham Hoover and Carol Griswold Hoover \\
Ruby Valley Research Institute                  \\
Highway Contract 60, Box 601                    \\
Ruby Valley, NV 89833                           \\
}

\date{\today}


\begin{abstract}
Following Berni Alder[ 1 ] and Francis Ree[ 2 ], Douglas Henderson was the third of Bill's California coworkers
from the 1960s to die in 2020\cite{b1,b2}. Motivated by Doug's death we undertook better to understand Lyapunov instability
and the breaking of time symmetry in continuum and atomistic simulations. Here we have chosen to extend our
explorations of an interesting pair of nonequilibrium systems, the steady shockwave and the unsteady rarefaction
wave. We eliminate the need for boundary potentials by simulating the collisions of pairs of mirror-images
projectiles. The resulting shock and rarefaction structures are respectively the results of the compression and
the expansion of simple fluids. Shockwaves resulting from compression have a steady structure while the rarefaction
fans resulting from free expansions continually broaden. We model these processes using classical molecular dynamics
and Eulerian fluid mechanics in two dimensions. Although molecular dynamics is time-reversible the reversed
simulation of a steady shockwave compression soon results in an unsteady rarefaction fan, violating the microscopic
time symmetry of the motion equations but in agreement with the predictions of macroscopic Navier-Stokes fluid
mechanics. The explanations for these results are an interesting combination of two (irreversible) instabilities,
Lyapunov and Navier-Stokes.

\end{abstract}

\maketitle

\section{Douglas James Henderson [1934-2020]}

Doug {\bf (Figure 1)} had the good fortune to do his ``Theory of Liquids'' PhD work with Henry Eyring at the
University of Utah in the company of Francis Ree. Francis completed his own PhD work on ``Random Walks''
with Eyring a year earlier.  Meanwhile Bill was completing his ``Virial Series'' PhD with
Andy De Rocco at the University of Michigan. All three budding chemical physicists, Doug, Francis,
and Bill, got together repeatedly over the next forty years. The details of Doug's research career
are spelled out in a 59-page {\it Curriculum Vitae} at Brigham Young University's ``Emeritus Faculty''
portion of the Chemistry and Biochemistry website.

After nearly a decade of teaching jobs at the University of Utah, Arizona State University,
and the University of Waterloo, Doug settled into a 20-year research career at IBM's Almaden
Research Center in San Jos\'e California.  It took only a little time for him to contact
like-minded physicists at the nearby Lawrence Radiation Laboratory. At LRL about a dozen such,
including Francis and Bill, had been brought together by the efforts of Berni Alder. All of us
had the benefit of working with Doug's colleagues at IBM, John Barker and Fareed Abraham, both of
them in the forefront of theoretical and computational explorations of fluid properties.

This IBM-LRL cross-fertilization led to joint work\cite{b3}, and to a series of West Coast
Statistical Mechanics Meetings, held at various University of California locations and the Almaden
Center. These activities continued through the personnel cutbacks of the early 1970s up to the 17 October 1989
Loma Prieta earthquake, which killed dozens and injured thousands of Bay Area residents. As a
consequence of the earthquake Doug relocated to the Latter-Day Saints' Brigham Young University at
Provo Utah, 600 miles east, over 800 by road.  A majority of his 500 publications were generated with
hundreds of colleagues all over the world following his 1990 move. His collaborations and travels
continued worldwide until his final illness this past Summer and Autumn of 2020.

Doug's commitment to his Latter Day Saints' faith was strong\cite{b4}.  In our experience most scientists'
faith, if any at all, makes up only a small portion of their day-to-day activities. By way of explaining this
Feynman gave a 1956 talk noting that science progresses through scepticism and doubt\cite{b5}.  Scientists
try to reduce the uncertainty inherent in models of the real world by comparing theories, experiments, and
computation. Feynman thought it difficult to accept the simultaneous realities of religious faith and
scientific discovery because any ``metaphysical'' [miraculous] aspects of faith, if definite enough to be
checked, might later be found to contradict scientific inquiry. Feynman's doubt that religious metaphysics
can withstand scientific investigation brings the example of evolution to mind. Doug's belief was clear\cite{b4}:
``Life is [not] possible without God''. Many, probably most, nonspecialist scientists would instead accept
Richard Dawkins' Darwin-based explanations as reasonable\cite{b6}.

\section{Doug's Scientific Work on Liquids}

Doug's seminal work with John Barker [1925-1995] on the understanding of liquid properties from a
knowledge of hard-sphere structural and thermodynamic properties was the product of decades of research.
We can do no better here than to quote from a 1998 review \cite{b7} of their joint contribution to the
Reviews of Modern Physics\cite{b8}, ``What is Liquid?'' :
\begin{quote}
it was a lifelong fascination: finding the interatomic and intermolecular forces,
models, and simulation techniques which could reproduce first qualitatively, but
ultimately quantitatively, liquid properties. Condensed rare gases were a recurring
subject in [their] studies. [Their] work spanned the transition from crude mechanical
liquid-structure models (Bernal's aggregates of steel ball bearings and Hildebrand's
coloured gelatin spheres, suspended in a fish tank, come to mind) to the computer
simulations of liquid structure which are commonplace today [1998]. In their early
work on liquid perturbation theory, Barker and Henderson discovered that the van der
Waals hard-sphere picture provided a workable structural model for computing static
liquid properties. Liquid free energies, suitable for thermodynamic calculations,
could be estimated by finding the density and temperature dependence of the optimum
hard sphere diameter.
\end{quote}

Onward from the 1970s hard-sphere perturbation theory has provided a useful route to the thermodynamic 
properties of simple liquids. By now this approach, for which Google returns over six million results,
has been superceded by computer simulation. Today it is feasible to compute hundreds, or even thousands,
of thermodynamic state points with straightforward Monte Carlo or molecular dynamics simulations. Pressure
and temperature, rather than energy and volume, can be made to serve as the independent variables for
atomistic simulations by using Nos\'e-Hoover dynamics, which was developed in the 1980s\cite{b9,b10,b11}.

Nonequilibrium versions of thermostated dynamics have provided accurate estimates of transport
coefficients, the shear and bulk viscosities along with the thermal conductivity. These linear transport
coefficients can also be obtained from equilibrium Green-Kubo correlation functions so that both the
equilibrium and hydrodynamic constitutive relations follow from relatively simple algorithms.
Wikipedia articles and Amazon's book offerings provide up-to-date expositions of the details. Through
the work of Doug and others the understanding of equilibrium thermodynamic properties is no longer a
mystery.  Today it is nonequilibium systems which provide interesting and challenging research areas.
They include the scientific subject of this memoir, the tension between the irreversibility of real
life and the idealized time-reversible models with which we describe irreversible processes.

\section{The Second Law and Time Reversibility}

``Time marches on''.  The steady, unstoppable advancement of time is all too apparent in writing a Memoir.
Rudolf Clausius recognized it in his famous 1865 version of the thermodynamic laws:
\begin{center}
``Die Energie der Welt ist constant.'' [ First Law ] \\
``Die Entropie der Welt strebt [strives] einem Maximum zu.'' [Second Law],
\end{center}
The Second Law describes the unstoppable aspect of real life, stating that an isolated system's entropy increases
to its equilibrium maximum.  In support of that law Ludwig Boltzmann's H Theorem showed that a probabilistic
description of dilute-gas collisions in an isolated system provides an (approximate) entropy model which cannot
decrease \cite{b12}. Thus it appears that this unidirectional law rules out the time-reversible nature of atomistic
mechanical models, both classical and quantum. The classical model leads to Liouville's [exact] Theorem,
which deals with an entropy defined by Gibbs' phase-volume formulation. For an isolated system the Theorem shows
that this entropy cannot change with time.

In principle,
but definitely not in practice, given initial conditions, the entire past and present of a Hamiltonian
system can be obtained by solving the motion equations. This oversimplified picture--the very existence of
longtime solutions--is subject to a serious caveat: in principle it is necessary to specify coordinates and
momenta with infinitely fine precision in order to begin the work of an ``exact'' solution.  No matter how
precise these conditions exponential Lyapunov instability guarantees an ultimate frustration.  In view of,
or even despite, this nit picking, it is certainly reasonable to wonder why time-reversible dynamics is such
a good model for the observed unidirectional evolution of the world around us. In our view this interesting
subject remains ripe for a better understanding based on manybody computer simulations of natural processes.
The inelastic collision of two crystallites or drops to make a single larger drop is an example\cite{b1,b13,b14}.
The organized motion required to reverse such a process looks entirely implausible. No doubt it is. On a more
macroscopic scale the spreading of Rayleigh waves on the surface of a pond, launched by a thrown rock, is another
such example. Reversing a strongly irreversible process requires an inordinate amount of information.

Conventional Navier-Stokes fluid mechanics supports the law. The instability of a violation, a reversed flow, is
built into the model. When the pressure-tensor component $P_{xx}$ includes the viscous stress $\eta(\partial u/\partial x)$
(with $\eta$ the shear viscosity) the Navier-Stokes motion equation is intrinsically irreversible. In a reversed
Navier-Stokes flow viscous stresses, proportional to the velocity gradient $(\partial u/\partial x)$ change sign,
leading to exponential instability. This instability is the result of poor modeling rather than Lyapunov's amplified
roundoff error.

Simulations can be decisive in understanding instabilities. {\bf Figure 2} is an illustration, showing both the stable
progress of a steady shockwave and the irreversibility of its reversed motion in a manageable small simulation of
8192 two-dimensional particles, a simulation requiring only a few minutes of computer time.

\section{The Role of Simulations in Understanding}

In our experience irreversible processes, when simulated numerically,
invariably do evolve in the direction of increasing entropy, even though entropy resists a useful definition for
nonequilibrium states.  Simulations provide the best means for understanding the details enabling the Second Law.
In {\bf Figure 2}, with the advantage of considerable simulations with varying velocities, timesteps, initial
conditions, and numbers of particles we adopted a short-ranged purely-repulsive pair potential for the dynamics\cite{b15} :
$$
\phi(r<1) = (10/\pi)(1-r)^3 \rightarrow \int_0^1 2\pi r \phi(r)dr \equiv 1 \ .
$$
Initially, in the uppermost frame, the entire 8192-particle cold crystallite of {\bf Figure 2} travels to the right at
velocity $(0.97,0.00)$. This choice of initial velocity yields the uniform equilibrium hot fluid shown in the third
configuration counting down from the top.  The hot fluid is now twice as dense as the initial cold crystallite, with the
conversion complete at a time $t = 114.6$.  Evidently the velocity of the left-moving shockwave is
$(-221.7025/2)/114.6 \simeq -0.97$, just as is required for a homogeneous twofold compression. The molecular dynamics
simulation in the Figure was continued onward for an additional time of 57.3, to the configurational snapshot at the
bottom of {\bf Figure 2}. Rather than retracing its recent history (as might well be expected from the symmetry of the
motion equations) the hot fluid generates something entirely new, the rarefaction fan shown at the bottom.

There are three different plausible futures from the maximum-compression state at time 114.6
shown in {\bf Figure 2}.  All of them, at time 171.9, are shown in {\bf Figure 3}: [1] at the top the straightforward
continuation of the molecular dynamics;
[2] in the middle the evolution to that same time, 171.9, with a Maxwell-Boltzmann distribution of velocities at the ``hot'' kinetic
temperature of 0.115; [3] at the bottom the result of changing the signs of all the velocities at the maximum-density
configuration at $t = 114.6$.  The failure of this last operation to recapture the initial condition at the top of {\bf Figure 2} illustrates
the Lyapunov instability of the reversed shockwave. This last future is not accessible to continuum mechanics due to that mechanicsÕ
lack of time reversibility.

By now both atomistic and continuum simulations of shockwaves are familiar. Our own work goes back to 1967. For a few
references see Bill's 1979 work\cite{b16} comparing Klimenko and Dremin's atomistic shockwave profiles to the predictions of
Navier-Stokes hydrodynamics. Back in 1967 Berni Alder had termed shockwaves the ``most-irreversible'' of processes\cite{b17},
suggesting that they are well worth studying. We continue that irreversibility work here. In the atomistic snapshots displayed in {\bf Figure 2}
it is noteworthy that the underlying equations of motion are Hamilton's, equivalent to Newton's $\{ \ F = ma \ \}$. Both mechanics
models are precisely time-reversible. Suppose we play a movie of the simulated shock process backwards.  What would we see? An
unstable impossible flow. The visual reversal is only made possible by previously storing the forward flow. In the projectile
simulation of {\bf Figure 2} we found that the largest Lyapunov exponent is of order unity in either time direction.  The exponent
describes the exponential growth rate of small perturbations. Thus the exponential growth of roundoff errors,
of order $e^{115}\simeq 10^{50}$ for the time interval illustrated in {\bf Figure 2}, is a Herculean barrier to longtime
reversibility.

The shockwave stagnation of this figure is taken from our recent Memoir to Berni Alder\cite{b1}. In this computational model, entropy
has clearly changed in the melting and heating of the once-cold projectile. Despite the time-reversibility of the underlying dynamical
simulation the second law of thermodynamics, entropy increase, is satisfied. As shocks are irreversible we were curious what would
happen when an atomistic reversible simulation was reversed, either by changing the signs of all the velocities or by changing the
sign of the timestep $dt$ in the integration algorithm.  What we found was surprising. A reversed steady shockwave, with a width of
about two particle diameters, was soon replaced by a nonsteady wave, growing in width linearly in the time, and soon spanning all
densities between that of the hot shocked fluid and a zero-density gas. We soon learned that such a structure, well known in fluid
mechanics, is a ``rarefaction wave'' and that its expansive behavior is easily understood for simple model fluids\cite{b18}, as we
will soon see.

At the start of the present work we guessed that a detailed picture of the reversed-shock instability could be obtained
by defining and characterizing local one-particle Lyapunov exponents.  Repetition of our earlier simulations\cite{b1}, along with
analyses of single-particle Lyapunov exponents was disappointing. Though conventional Lyapunov analysis places the instability squarely
at the shock, single-particle analyses provided no useful information. We abandoned their study and took up the analyses of shocks
and rarefactions which makes up the balance of this paper.

\section{Four Implementations of Shockwave Simulations}

{\bf Figure 4} illustrates three methods for generating shockwaves with projectiles and boundary forces, as in our previous work.
At the bottom we illustrate a fourth approach, developed for the present work. It avoids boundary potentials by modeling the collision of
two mirror-image projectiles. Using the same short-ranged soft-disk potential model as before\cite{b15,b1}, with projectile velocities of
$\pm 0.97$ we obtained twofold compression of the original stress-free triangular lattice.  {\bf Figure 5} illustrates the collision of
two 4096-particle projectiles, with a fourth-order Runge-Kutta timestep of 0.01, and an initial kinetic temperature (to break the
perfect-lattice symmetry) of 0.0001. This simulation shows that the stress-free cold projectiles at density $\sqrt{(4/3)}$ and temperature
0.0001, are steadily converted to hot fluid, the same fluid state, with density $2\sqrt{(4/3)}$ and kinetic temperature 0.115, as in our
earlier work\cite{b1}, but without the need of any boundary forces.

Because two shocks rather than one are involved in the conversion the projectile method requires only half the time of our previous
stagnation-based work. Both projectiles are twofold compressed at a time of 58. {\bf Figure 6} shows the steady decline of the total
kinetic energy $K$ during compression, followed by its recovery during the rarefaction [expansion] which begins at maximum compression.
It is noteworthy that the time variation is symmetric about the minimum within an accuracy of about one percent.

 {\bf Figure 5} shows five snapshots of the 8192-particle
two-projectile  system, equally spaced in time with an interval of 29.  The middle image is accurately compressed twofold, but its
subsequent expansion is not at all like that of its past history.  The pair of rarefaction waves visible in the fourth and fifth images
were a complete surprise. Let us focus next on what happens to the hot twofold-compressed fluid in the center of the figure when it expands.
In order to analyze the results we introduce smooth-particle averaging, a useful technique for generating continuously differentiable
field variables from discrete atomistic information.

\section{Smooth-Particle Averaging of Density Profiles}

Just as in the single-projectile simulations of {\bf Figures 2 and 3}, we consider three different continuations of molecular dynamics
simulations, in {\bf Figure 7} from the middle view of {\bf Figure 5}: [1] Conventional molecular dynamics; [2] Molecular dynamics with the
initial velocities chosen from an equilibrium Maxwell-Boltzmann distribution; [3] Molecular dynamics with all velocities reversed.
To make quantitative comparisons of the densities in these three different scenarios beginning with the maximum-density configuration we will
evaluate a local smoothly-continuous density function, $\rho(x)$ for each of them. We use a ``smooth-particle'' algorithm based on Lucy's
weight function\cite{b1} with a smoothing length of $h=3$. The weight function $w(|dx|)$ is zero for distances beyond 3 and has two vanishing
space derivatives there, guaranteeing a smoothly differentiable density profile throughout its range:
$$
w_{\rm Lucy} = (5/4hL_y)(1 - 6z^2 + 8z^3 - 3z^4) \ {\rm with} \ z = |dx|/h \ \longrightarrow
$$                                                                                                                                                  
$$
\begin{textstyle}
\int_{-L_y/2}^{+L_y/2}dy\int_{-L_x/2}^{+L_x/2}dx \ w_{\rm Lucy}(|x-x_g|) \equiv 1 \ {\rm for} \ |x_g| < (L_x/2)-h \ .
\end{textstyle}
$$
With 8192 particles and original projectile dimensions $[L_x=256\sqrt{(3/4)}=221.70]\times [L_y = 32]$, densities in
the range $|x_g| < (L_x/2)-h$ represent averaged contributions from an area of $32\times 6 = 192$. The density at any
such gridpoint location $x_g$ is simply given by the sum of all such nearby particles' smoothed contributions,
$$
\rho(x_g) \equiv \sum_N w_{\rm Lucy}(|x-x_g|<3) \ .
$$
{\bf Figure 8} compares the three future density profiles: molecular dynamics, Maxwell-Boltzmann, and reversal. The
detailed evolution appears in {\bf Figure 9} for the two-projectile method. The cold material density, $\sqrt{(4/3)}
= 1.1547$, is doubled to 2.3094 in the shocked fluid.

Continuing the simulation over the interval $58 < t < 116$ with either the old velocities or with new ones from a $T=0.115$
Maxwell-Boltzmann distribution makes no significant difference.  We conclude from this comparison that the maximum-compression
configuration is effectively at equilibrium and that reversing the velocities indicates a nonequilibrium correlation between
coordinates and velocities that recaptures a part, but far from all, of the original time-zero configuration. Evidently the
molecular dynamics, due to Lyapunov instability of order $e^t$, is unable to recapture the past. No doubt that a sufficiently
precise simulation (well beyond research budgets) could reproduce the initial configuration within visual accuracy. Trials using
timesteps $\{ \ 0.04, 0.01, 0.0025, 0.000625 \ \}$ as well as the (orders of magnitude slower) quadruple precision version of the
Runge-Kutta integrator confirmed that storing the old configurations or using a bit-reversible algorithm are the only feasible
approaches to capturing the past. Recently we considered the simulation of a similar single-projectile rarefaction problem\cite{b1}.
We found that reversing the velocities at maximum compression gave a largest Lyapunov exponent with its largest contributors
localized near the shockwave front.  That finding indicates that the shockwave itself is the main barrier to time reversibility.

\section{Hydrodynamic Solutions with Centered  Differences}

The steady nature of shockwaves is well known\cite{b18} and leads to a simple numerical method for finding their structure
from a known model for the constitutive equations.  Here we choose simple model mechanical and thermal equations
of state suggested by linear dependences of the energy on density and temperature and satisfying the Rankine-Hugoniot
relation equating the work done in shock compression to the energy change:
$$
P = \rho e \ ; \ e = (\rho/2) + T \ ; \ e_H - e_C \equiv (1/2)(P_H+P_C)(V_C-V_H) \ .
$$
This relation follows from the ``Symmetric'' compression picture from {\bf Figure 4}.
Choices of pairs of mechanical and thermal variables satisfying the Rankine-Hugoniot relation for the model include
$$
\rho: 1 \rightarrow 2 \ ; \ u: 2 \rightarrow 1 \ ; \ P: (1/2) \rightarrow (5/2) \ ; \ T: 0 \rightarrow (1/4) \ ; \
e: (1/2) \rightarrow (5/4) \ .
$$
To complete the constitutive equations needed for continuum simulations we choose to use a constant unit shear
viscosity, setting the scale of the model's shock width at unity.  The pressure tensor is $P_{xx} = P - (du/dx) \ ;
 \ P_{yy} = P + (du/dx)$, where $P$ is the equilibrium pressure. For simplicity we ignore heat conductivity. The
steady nature of the wave implies that the mass, momentum, and energy fluxes are all constant. In the case we
have chosen here they are
$$
[ \ \rho u \ ] = 2  \ ; \ [ \ P_{xx}  + \rho u^2 \ ] = (9/2) \ ; \ [ \ (\rho u)(e + (P_{xx}/\rho) + (u^2/2)) \ ] = 6 \ .
$$
Knowing the values of the three constants the result can be written as an ordinary differential equation for the
velocity gradient, $(du/dx)$.  A Runge-Kutta solution for a similar problem\cite{b19} appears in {\bf Figure 10}. 

Another (more challenging) numerical approach\cite{b20} solves the partial differential evolution equations on a fixed
``Eulerian'' grid.
The resulting partial differential equations are as follows:
$$
(\partial \rho/\partial t) = -(\partial( \rho u)/\partial x) \ ; \
(\partial u/\partial t) = -u(\partial u/\partial x) - (1/\rho)(\partial P_{xx}/\partial x) \ .
$$
$$
(\partial e/\partial t) = -u(\partial e/\partial x) - (1/\rho)P_{xx}(\partial u/\partial x) \ .
$$
Notice that the viscous term in $P_{xx}$ contributes to a diffusion equation, $(\partial u/\partial t) \propto
(\partial^2 u/\partial x^2)$, leading to decay, $\propto e^{-x^2/4t}$, rather than divergent \
growth, $\propto e^{+x^2/4t}$, which would follow from changing the direction of time
.
We were able to find solutions by following the description given on page 360 of our {\it Kharagpur Lectures}
book\cite{b20}.  It is interesting that the time-dependent equations require a much smaller step than do the
steady state equations.  See {\bf Figure 12} for details. We verified that the two methods, steady-state and
Eulerian,  are consistent and turned to considering the detailed
differences between shock and rarefaction waves.

The usual textbook explanation of such a wave's behavior can be applied to the existence of shock and rarefaction
waves. In a typical fluid the sound velocity increases with density. Imagine the initial density distribution
shown in the caricature of {\bf Figure 11}.  A wave traveling rightward must spread while a wave traveling leftward
must steepen, reaching a finite width due to the diffusive nature of viscosity. The rightmoving wave's width grows
without bound in the absence of physical boundaries.
This simple explanation, illustrated in {\bf Figure 11}, is responsible for the extreme instability of the shockwave
to time reversal.  A molecular dynamics simulation with $dt = 0.000625$ keeps the energy constant to over a dozen
digits throughout a fourth-order Runge-Kutta simulation, but still, in a relatively short time the reversed
shockwave becomes a rarefaction.

{\bf Figure 12} illustrates the transformation, using 3000 spatial intervals and a timestep of 0.00005.  Even so we
found it necessary to remove an even/odd instability periodically (every 20,000 steps) by averaging the density,
velocity, and energy according to the following scheme\cite{b20}:
$$
f^{new}_i = [f^{old}_{i-1} + 2f^{old}_i + f^{old}_{i+1}]/4 \ .
$$
The reversed shockwave broadens roughly linearly in the time, as can be seen in the Figure.
With this successful simulation of the conversion of a shockwave to a rarefaction wave following velocity reversal
we see that our molecular dynamics simulations are quite consistent with the predictions of simple computational
fluid dynamics.

\section{Conclusions and Recommendations}

In this work we set out to avoid the flow perturbations introduced by boundary potentials and found that simulations
using mirror-image pairs of projectiles provide high-quality simulations of shock and rarefaction waves.  Smooth-particle
averaging makes it possible to compare atomistic and continuum profiles.  One can also formulate Lucy function
recipes for the local thermodynamic properties. Local averages of kinetic temperature and the density can provide local
entropies, from which entropy production could be analyzed. Although the future choices of Lyapunov-unstable dynamical
simulations must depend upon the local density of accessible phase-space states, accessible through roundoff error
and its amplification, we have not attempted such a project.  No doubt a simpler few-body nonequilibrium system,
such as a short conducting $\phi^4$ chain\cite{b21} or a single heat-conducting oscillator in a temperature 
gradient\cite{b22} would provide a better starting point.

The hydrodynamic shock/rarefaction instability we have discussed and modeled here is just one among the many
instabilities found with continuum constitutive relations. The Rayleigh-Taylor instability, with a denser fluid pressing
upon a less-dense one, is similar.  The reason why the shockwave/rarefaction instability discussed here has no classic
name is that laboratory experiments (as opposed to computer experiments) do not yet have the capability of reversing all
the velocities in a fluid system!

\section{Acknowledgment} We are grateful to the referee for his careful reading of
the manuscript.

\begin{figure}
\includegraphics[width=2.5 in,angle=-90.]{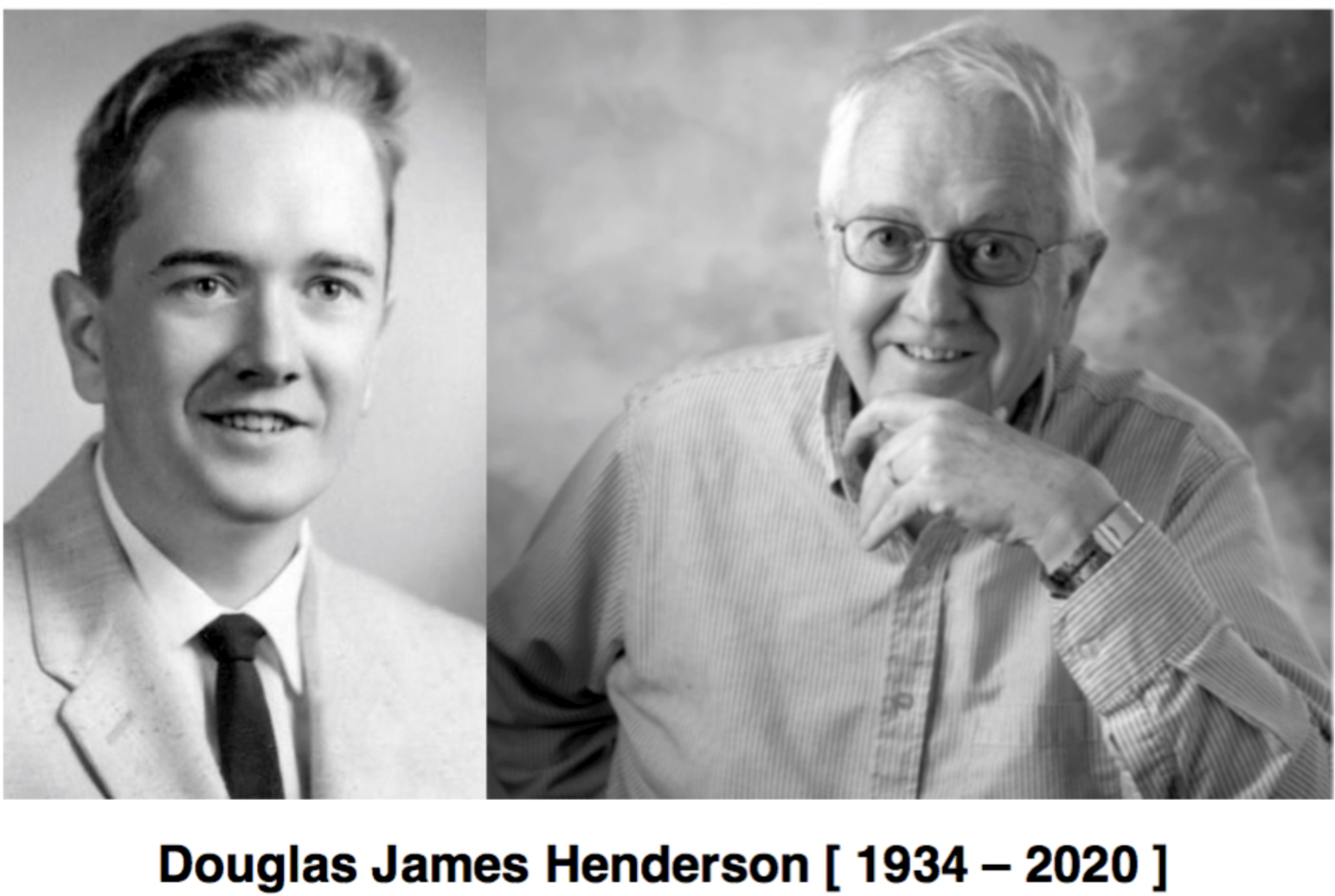}
\caption{
Douglas James Henderson [1934-2020]
}
\end{figure}

\begin{figure}
\includegraphics[width=2.5 in,angle=+90.]{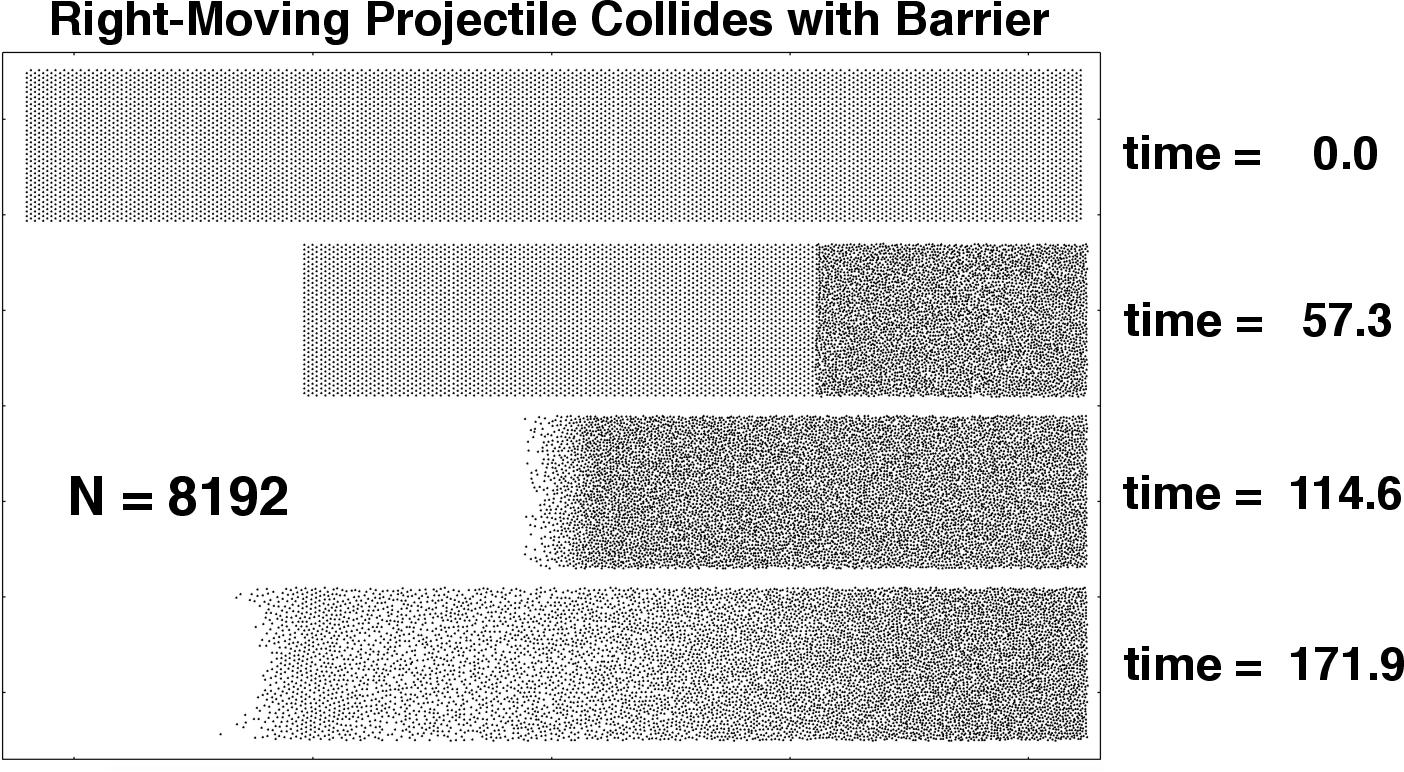}
\caption{
8192 particles with initial periodic height 32 and length $L_x = 256\sqrt{(3/4)} = 221.7025$ with a
fixed external quartic potential at $L_x/2$. The initial density is $\sqrt{(4/3)}$ with
velocity 0.97, leading to twofold compression and a final temperature of 0.115 at time 114.6. The
initial state has Maxwell-Boltzmann velocities centered on (0.97,0.00) with $T = 0.0001$ to break
the crystal symmetry. The final state at time 171.9 illustrates the irreversibility of a shockwave,
with the dynamics producing a rarefaction wave shown at the base of the figure.
}
\end{figure}

\begin{figure}
\includegraphics[width=3 in,angle=-90.]{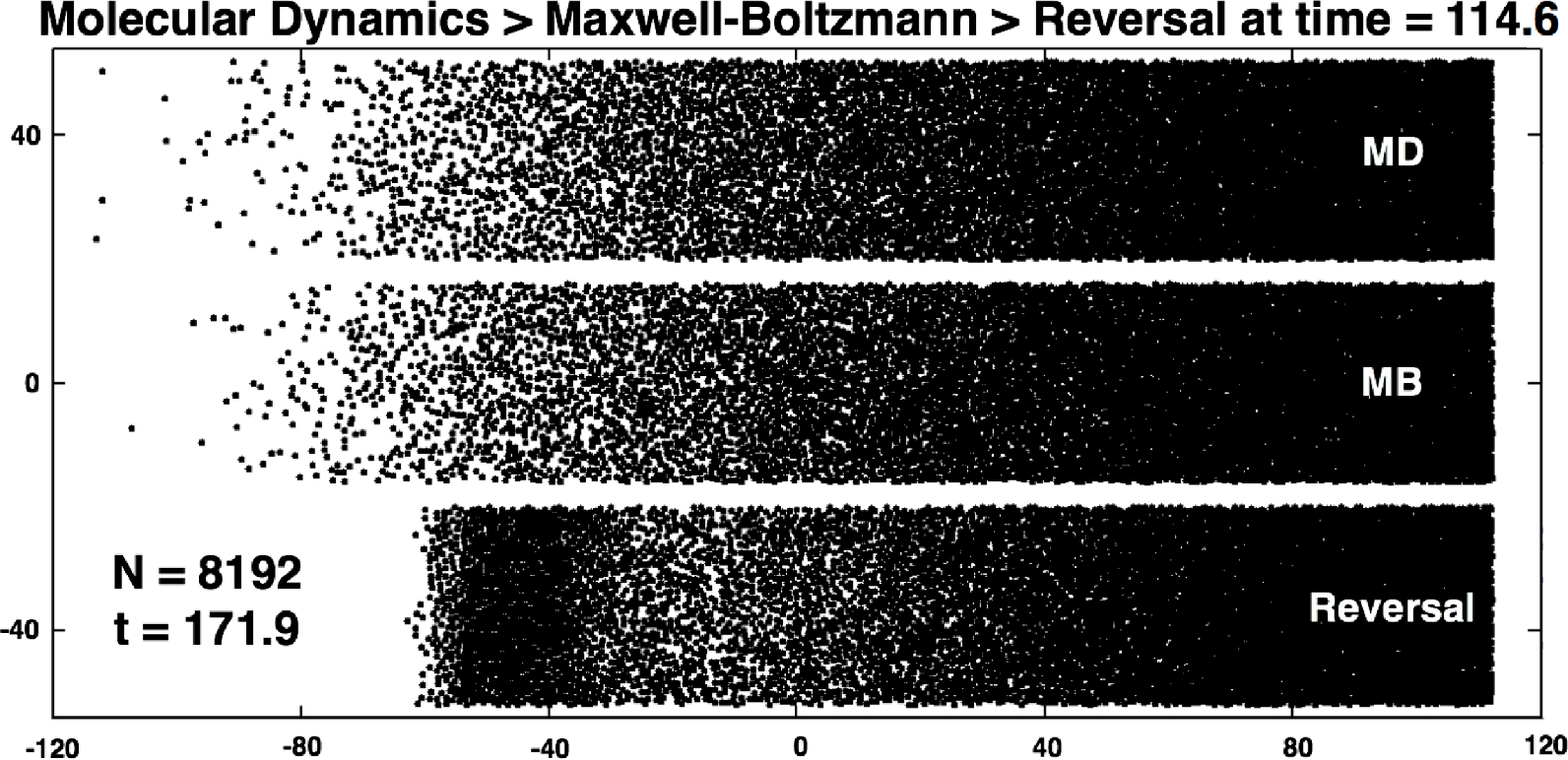}
\caption{
Three future scenarios for the third snapshot of Figure 2. There is no significant difference
between the configurations using molecular dynamics (at the top) or a Maxwell-Boltzmann
distribution at time 114.6 with a kinetic temperature of 0.115.  On the other hand the choice
of reversed velocities at 114.6 exhibits memory at the left of the earlier configuration at
time 57.3. Throughout this work the vertical boundary conditions are periodic.
}
\end{figure}

\begin{figure}
\includegraphics[width=3 in,angle=-90.]{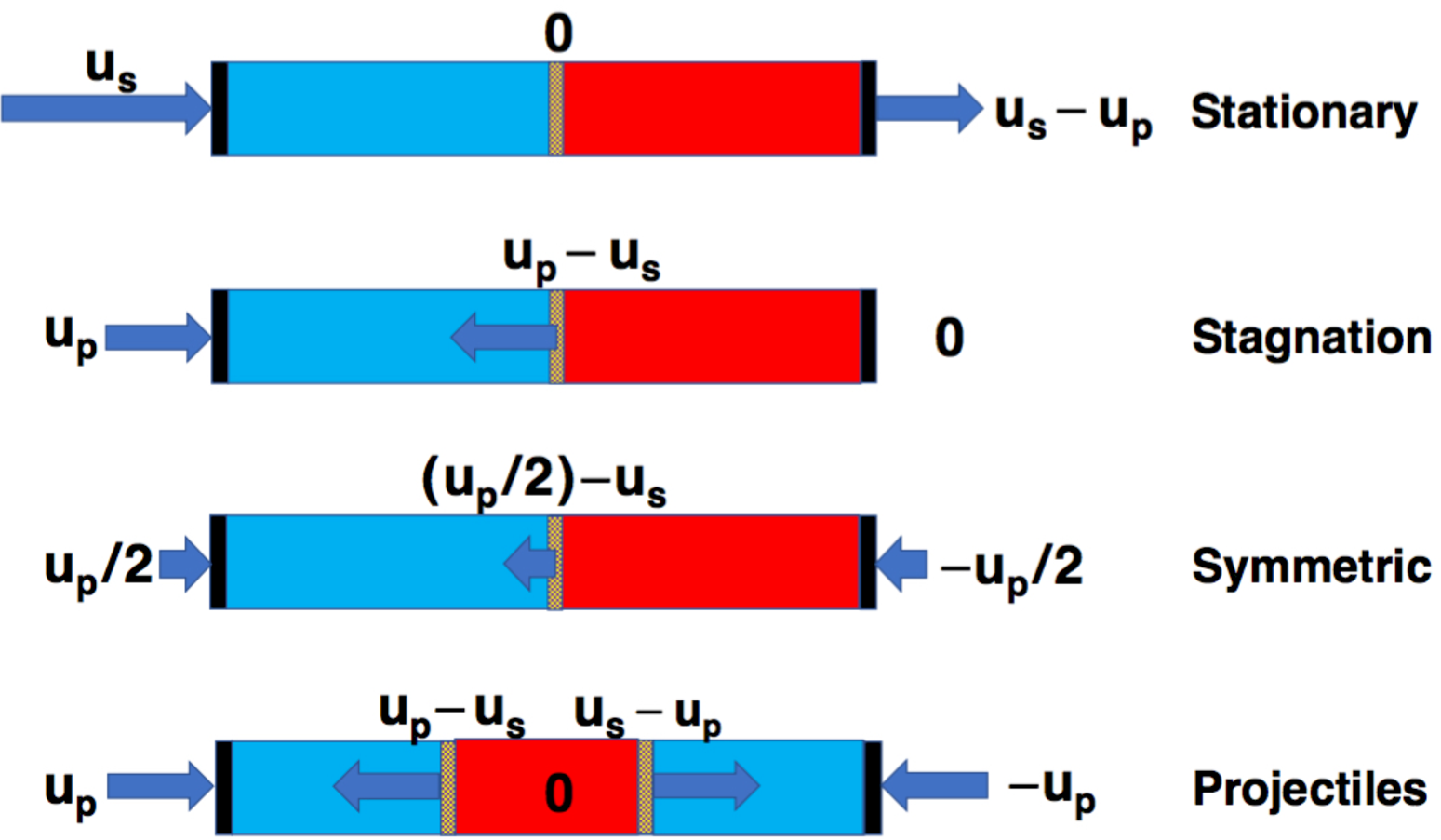}
\caption{
Four methods for the generation of shockwaves. Those waves, separating cold blue crystal from hot
red fluid are shown as yellow/black checkerboard.  The present work is based on a twin-projectiles
method, firing two mirror-image stress-free solids at one another, resulting in twofold compression
at a kinetic temperature of 0.115 and a density twice the original, $\rho_H = 2\sqrt{(4/3)} = 2.3094.$
}
\end{figure}

\begin{figure}
\includegraphics[width=3 in,angle=-90.]{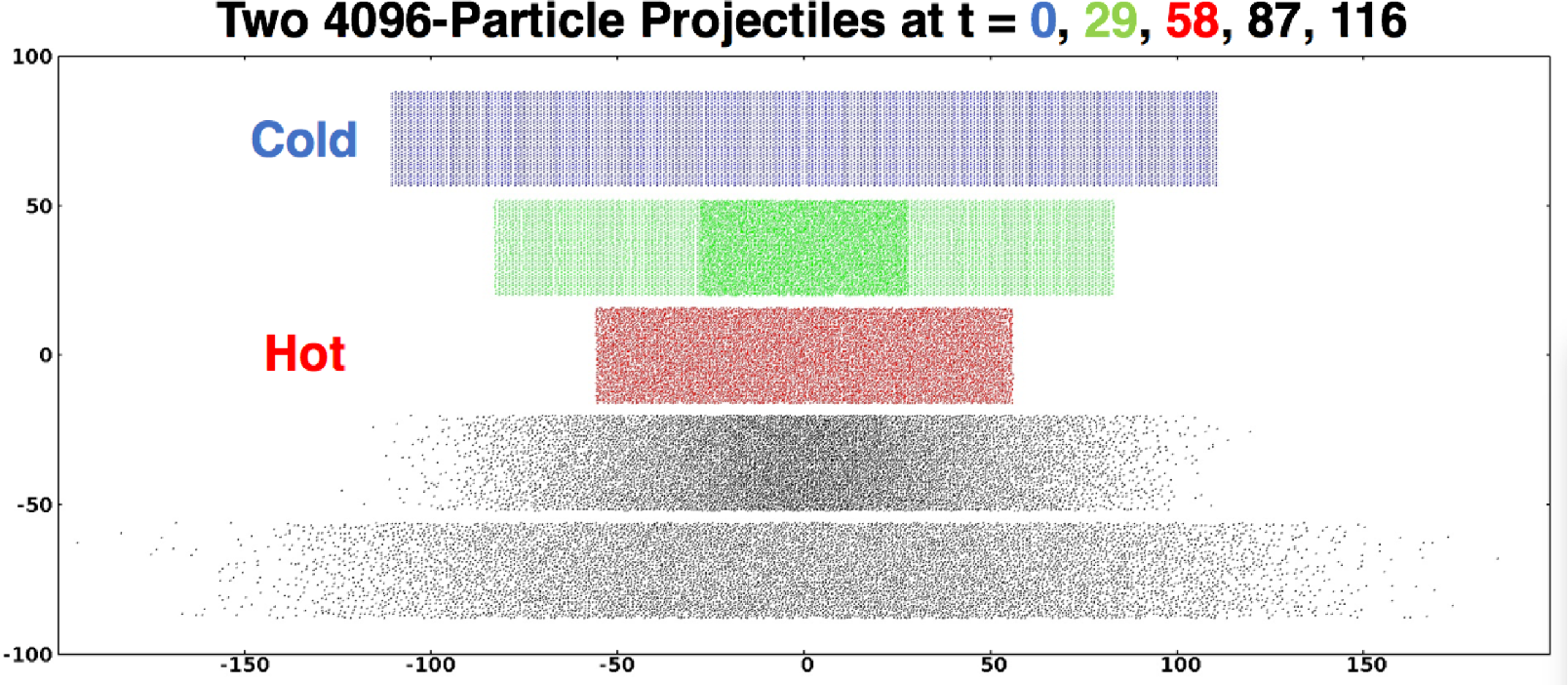}
\caption{
Collision of two 4096-particle projectiles at velocities $(\pm 0.97, \ 0.00)$ equally spaced in time.
Just as in the single-projectile simulations the maximum-density stagnant configuration gives
rise to rarefaction waves, two of them in this twin-projectile simulation. Vertical boundary conditions
are periodic.
}
\end{figure}

\begin{figure}
\includegraphics[width=3 in,angle=-90.]{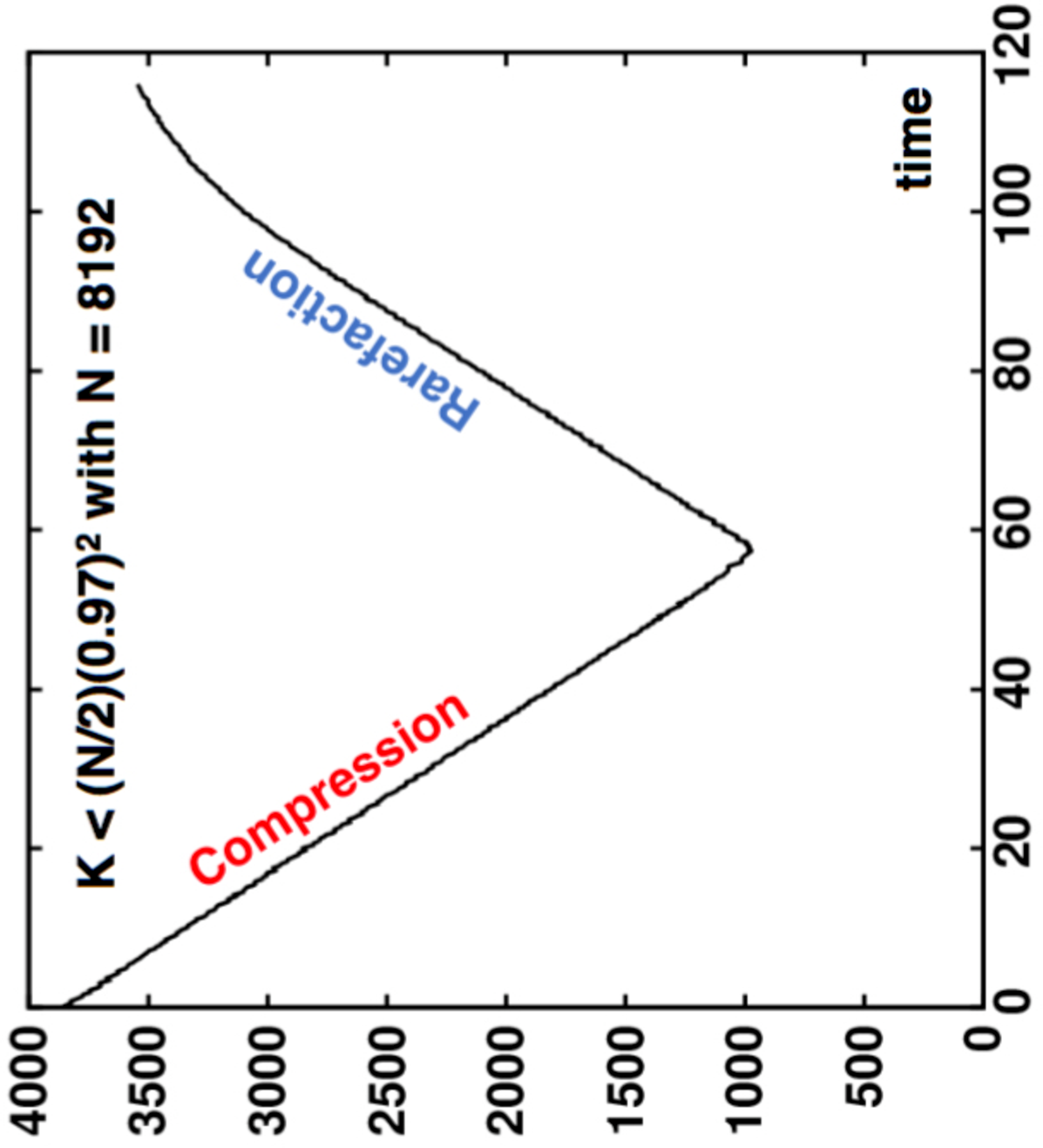}
\caption{
Kinetic Energy of two projectiles with standard molecular dynamics and $dt = 0.01$ showing
conversion of cold to hot material is complete at $t = 58$ . The kinetic temperature increases
from 0.0001 to 0.115 in the compression process and decreases steadily throughout the expansion
of the rarefaction fan.
}
\end{figure}

\begin{figure}
\includegraphics[width=2.1in,angle=-90.]{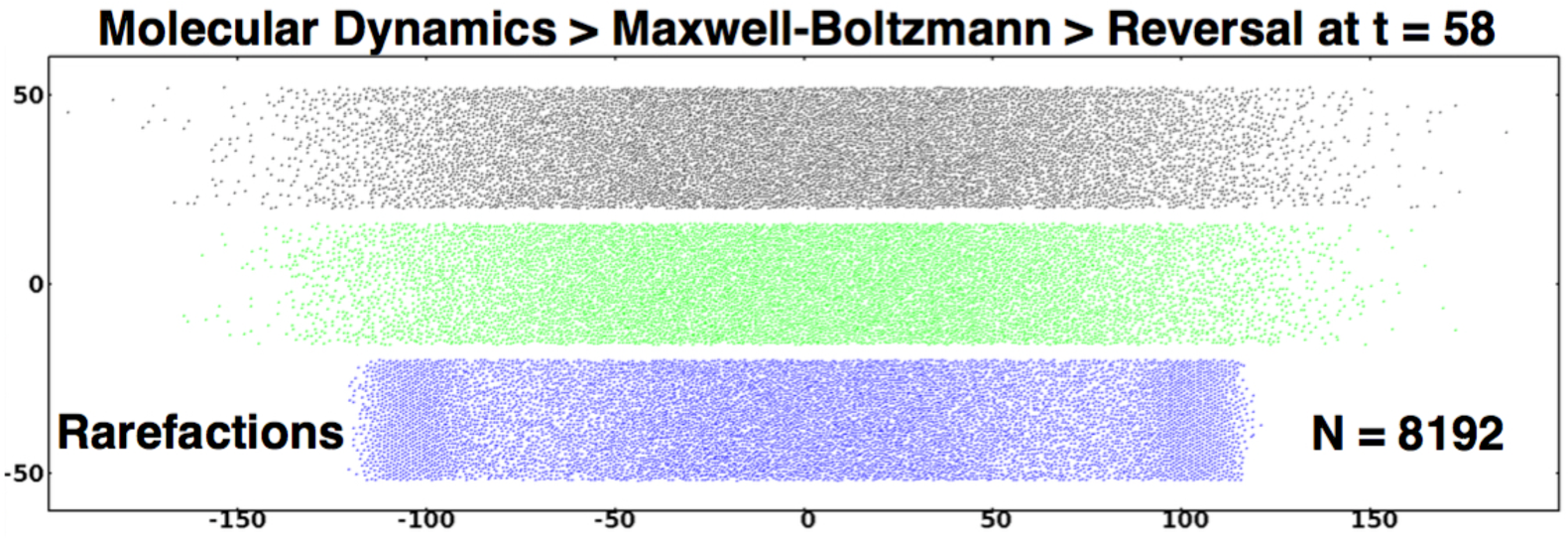}
\caption{
Two-projectile configurations at $t = 116$ using, top-to-bottom, standard molecular dynamics; replacing the
velocity distribution at time 58 with a Maxwell-Boltzmann distribution at kinetic temperature $T = 0.115$;
changing the sign of $dt$ (or all of the velocities) at $t = 116$. The vertical boundary conditions are periodic.
}
\end{figure}

\begin{figure}
\includegraphics[width=2 in,angle=-90.]{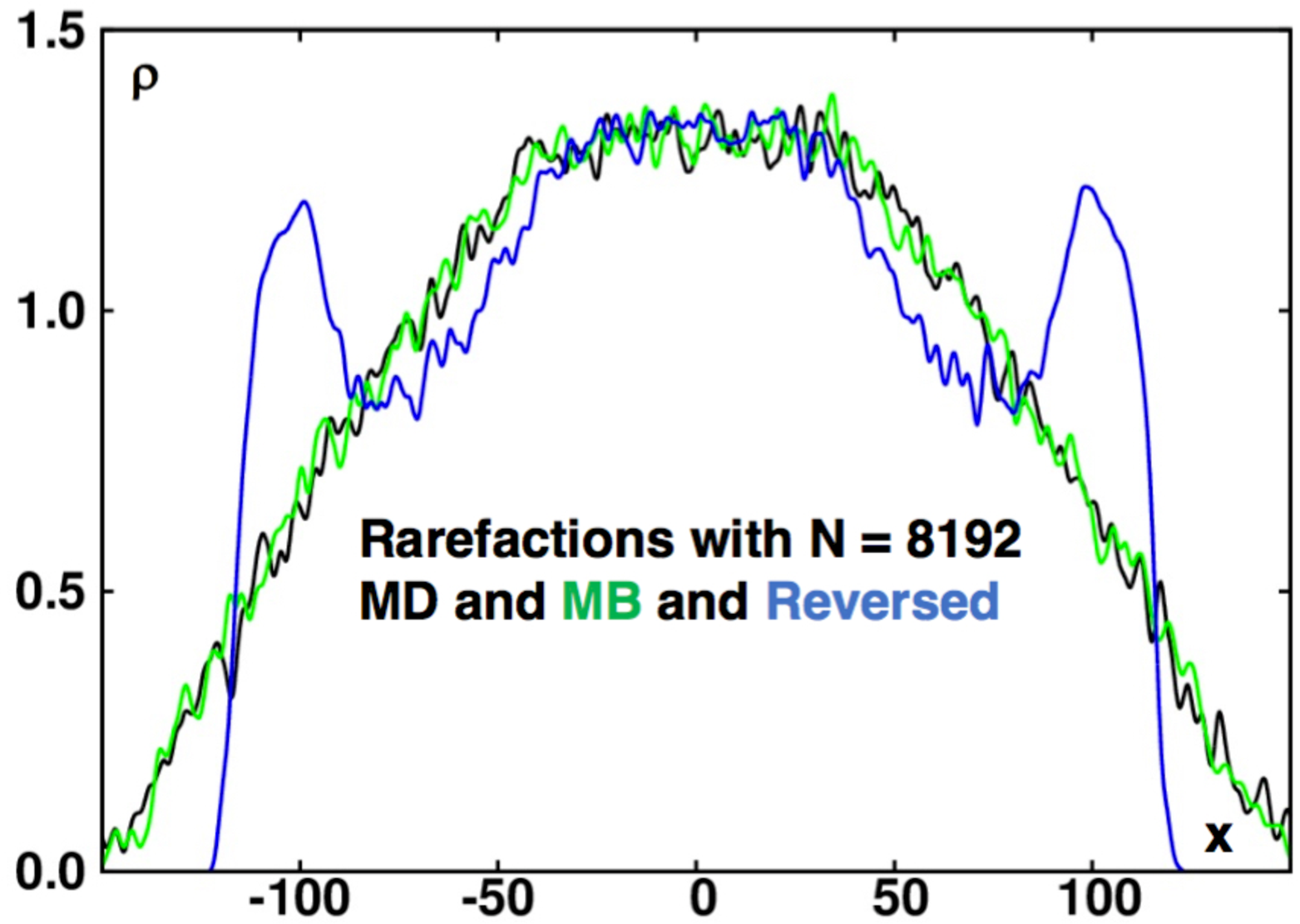}
\caption{
Density profiles using Lucy's smooth-particle weight function at time 116 where three distinct
differences were imposed at the time of maximum compression of two colliding projectiles, $t=58$.
The black profile results from standard molecular dynamics. The green profile results when a
Maxwell-Boltzmann velocity distribution with $T= 0.115$ is imposed on the maximum-compression
configuration at $t=58$. A precise time reversal of the molecular dynamics velocities at that time
results in the blue density distribution at time 116. In this last case a memory of the initial
time-zero density of $\sqrt{(4/3)} = 1.15470$ remains in the wings near $x \simeq \pm 120$. The
configuration corresponding to those sharp peaks can be seen at the bottom of {\bf Figure 7}. The
rarefaction fans generated with molecular dynamics and with Maxwell-Boltzmann velocities introduced
at maximum compression are statistically indistinguishable. 
}
\end{figure}

\begin{figure}
\includegraphics[width=3 in,angle=-90.]{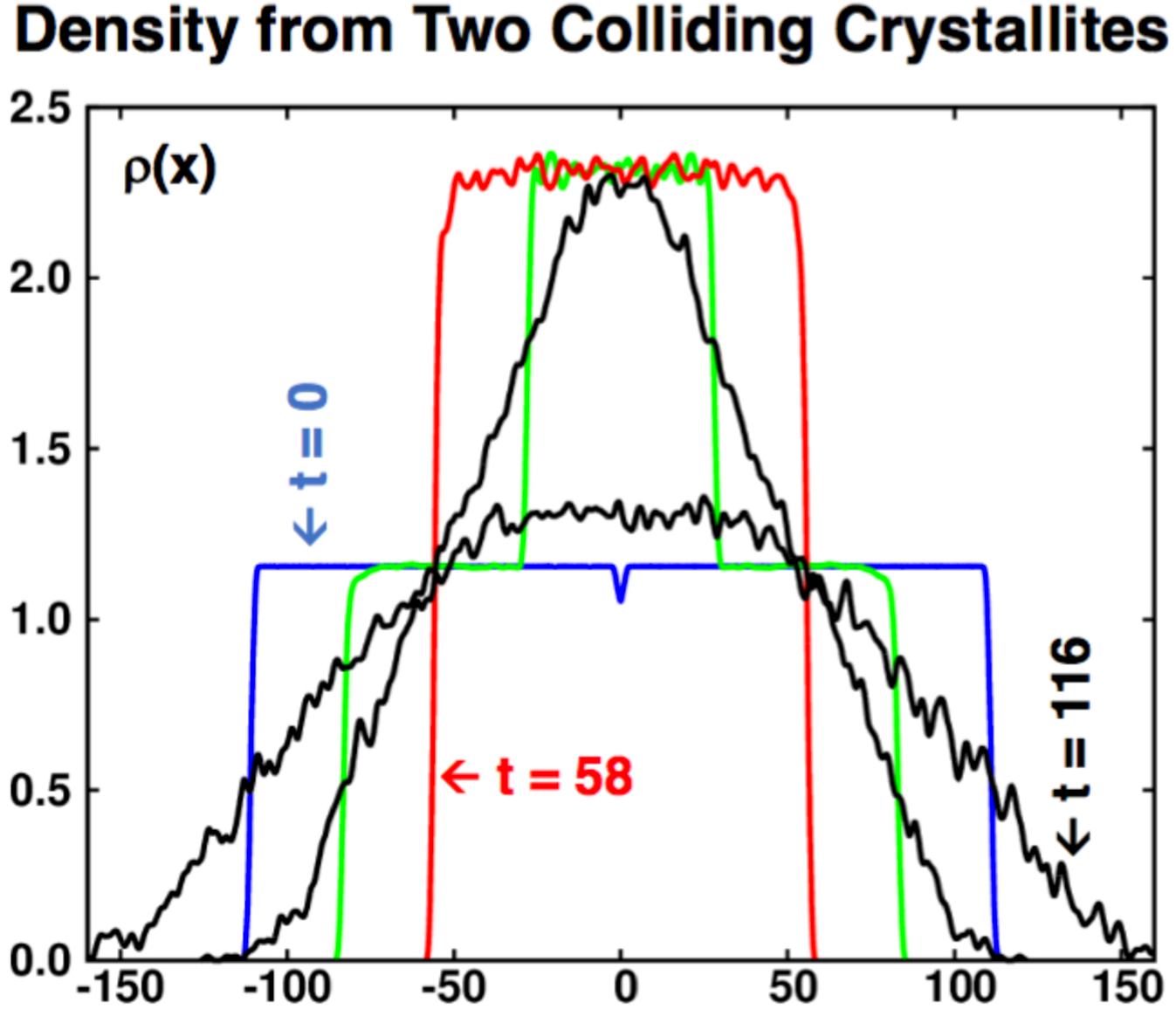}
\caption{
Lucy function densities from two colliding 4096-particle projectiles with a smoothing range $h=3$, timestep 0.01.
The projectile heights are 32. The lengths are $128\sqrt{(3/4)}$.
The initial projectile velocities, $\pm 0.97$ are converted into internal energy at the time of the third
snapshot, $t=58$. The rarefaction fan covers the whole density range in the fourth snapshot at time 87. There is no
tendency for the projectiles to separate.
}
\end{figure}

\begin{figure}
\includegraphics[width=3 in,angle=-0.]{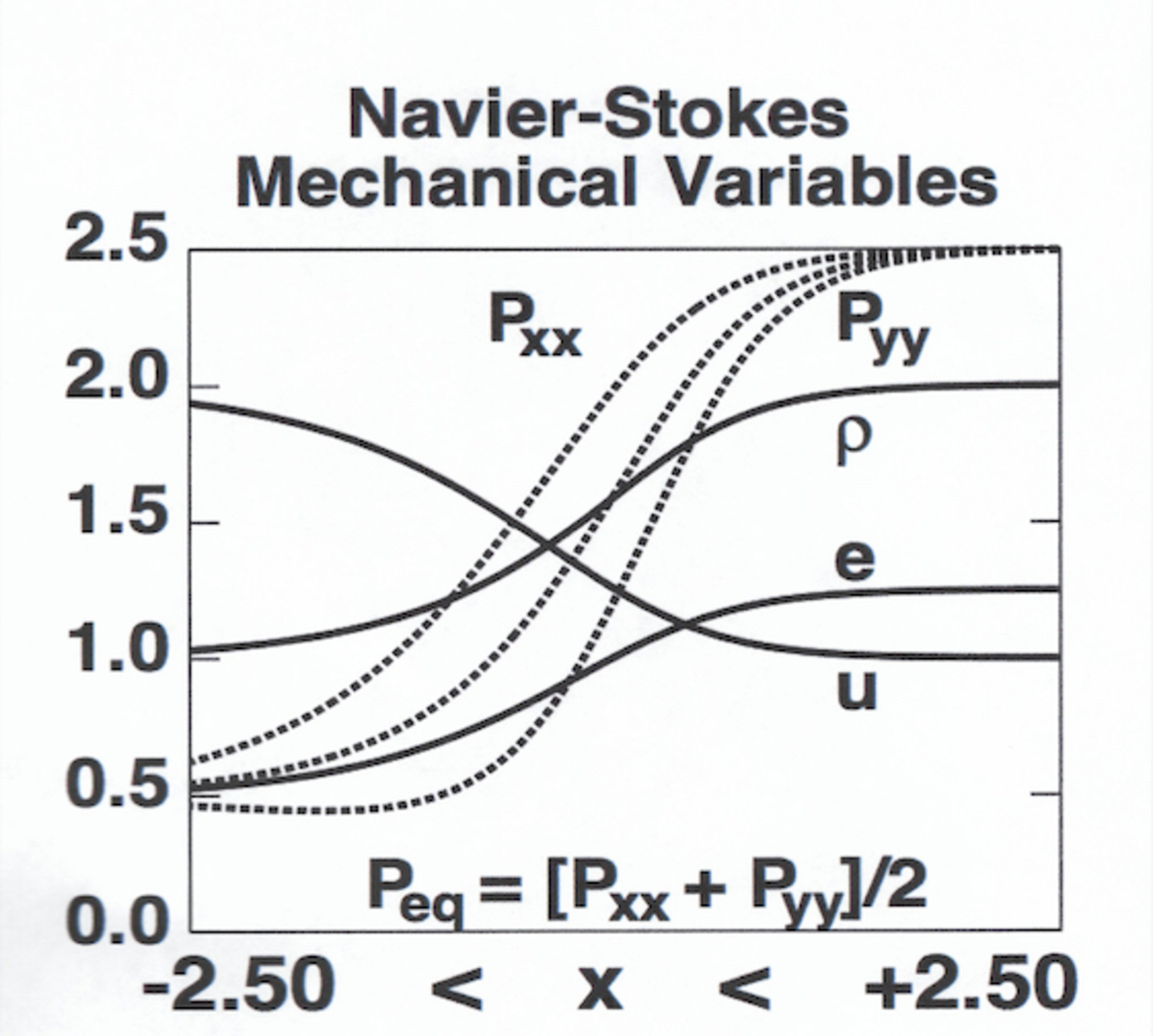}
\caption{
A stationary shockwave with mass, momentum, and energy fluxes of 2, (9/2), and 6 taken from page 18 of
reference 19. In the state obtained by reversing a movie of the wave the velocity $u$ changes sign so 
that the pressure tensor components $P_{xx}$ and $P_{yy}$ change places, corresponding to a negative shear
viscosity.
}
\end{figure}

\begin{figure}
\includegraphics[width=3 in,angle=-90.]{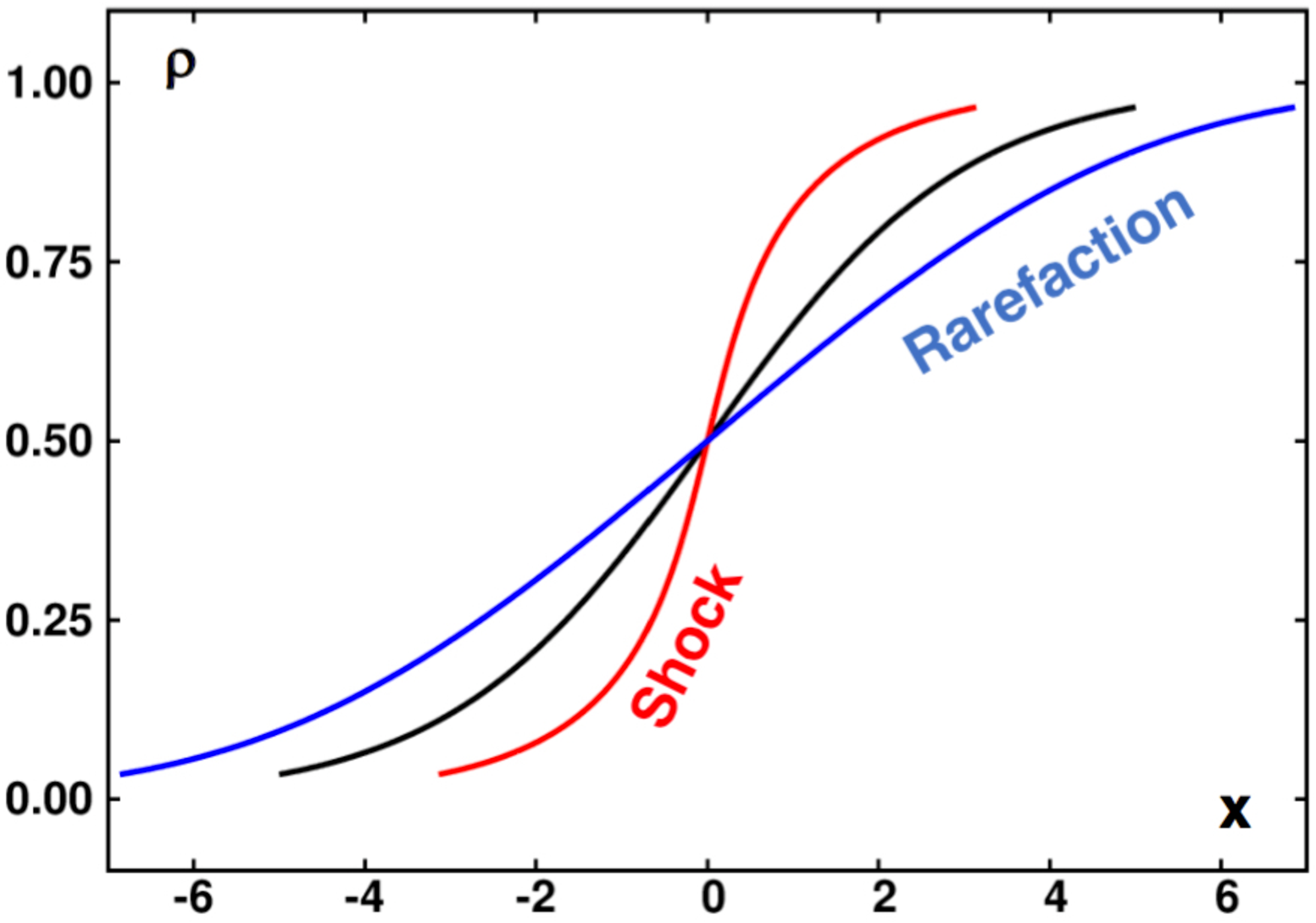}
\caption{
An initial wave (black) if moving to the right becomes rarefied (blue). If moving to the
left the increase of sound velocity with density generates a shockwave (red).  The finite
width of a shock is the result of viscosity. This irreversible argument is consistent with
the results of the reversible atomistic simulations of {\bf Figures 2 and 5}.
}
\end{figure}

\begin{figure}
\includegraphics[width=3 in,angle=-90.]{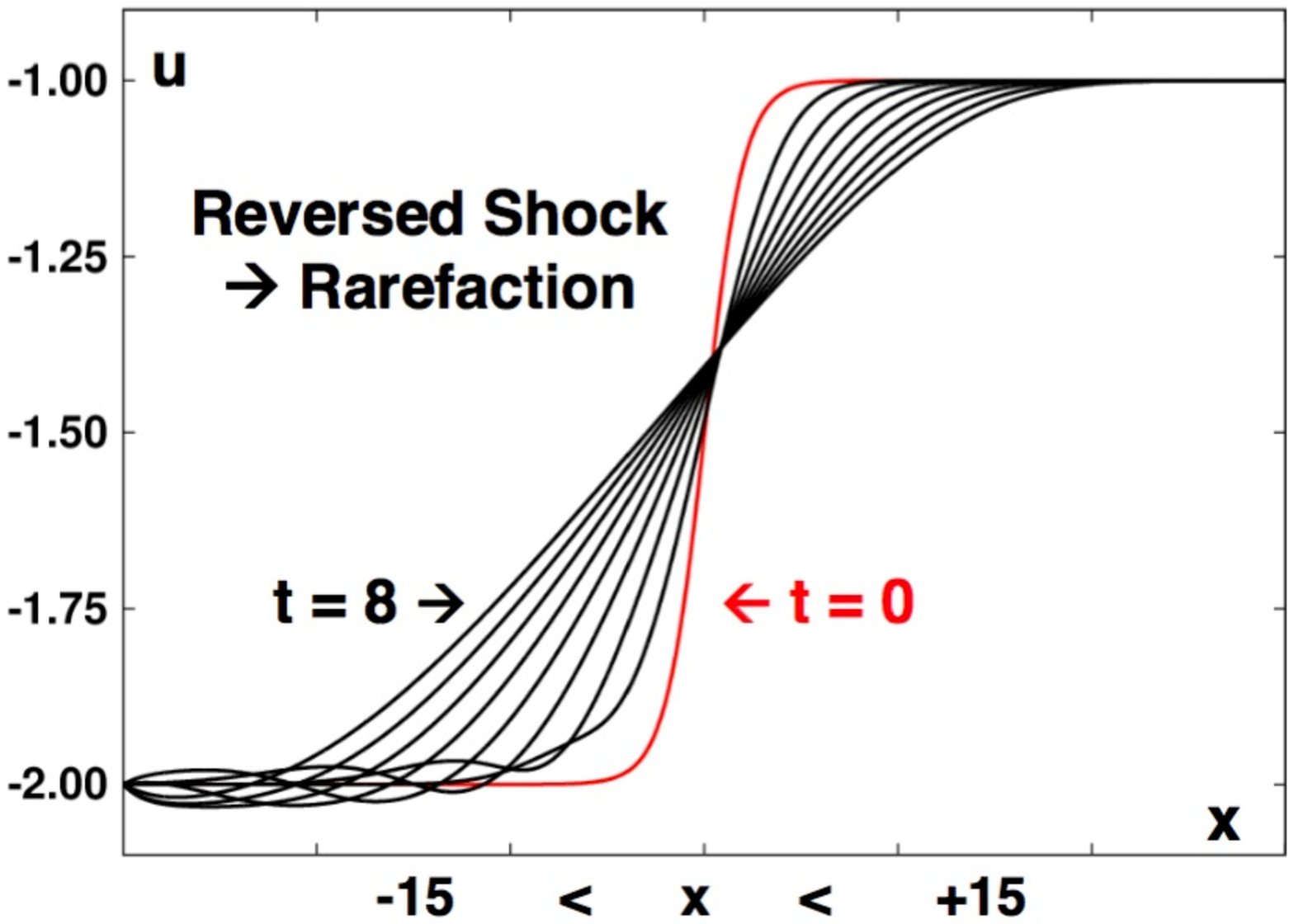}
\caption{
Reversal of the red steady-state Navier-Stokes shockwave velocity promptly results in a rarefaction wave,
snapshots of which are in black up to a time of 8. Centered finite-difference solution with timestep
0.00005 with even/odd instability removed every 20000 timesteps.
}
\end{figure}

\end{document}